# Exploring rotational resonance in elastic metamaterial plates to realize doubly negative property

Wei Wang[1], Bernard Bonello[1*], Bahram Djafari-Rouhani[2], Yan Pennec[2], and Jinfeng Zhao[3]

[1]Sorbonne Université, UPMC Université Paris 06 (INSP–UMR CNRS 7588),
4, place Jussieu 75005 Paris, France

[2]Institut d'Electronique, de Micro-électronique et de Nanotechnologie (IEMN–UMR CNRS 8520),
Université de Lille Sciences et Technologies, Cité Scientifique, 59652 Villeneuve d'Ascq Cedex, France

[3]School of Aerospace Engineering and Applied Mechanics, Tongji University, 100 Zhangwu Road,
200092 Shanghai, China

*corresponding author: bernard.bonello@insp.jussieu.fr

**Abstract**:

We report the realization of simultaneously negative effective mass density and shear modulus in a single-phase asymetric double-sided pillared metamaterial. The negative effective mass density is achieved by the combination of bending and compressional resonances of one pillar whereas the rotational resonance of the other pillar leads to the negative effective shear modulus. The coupling between these two pillars is investigated to describe the formation of the doubly negative property. Then, a pillared system featuring chirality is designed in order to make efficient the excitation of the rotational vibration, the occurrence of which is demonstrated by the transmission spectrum. Finally, numerical simulations of the zero-index refraction are carried out to prove the occurrence of the doubly negative property.

The advent of locally resonant metamaterials almost two decades ago,[1] and the great deal of research that ensued,[2–7] have significantly contributed to the possibilities we have now for controlling the propagation and the dispersion of acoustic/elastic waves. Some effective elastic properties of these artificial structures, may exhibit abnormal behaviors in narrow frequency bands where they may be infinite positive, null or even negative.[8,9] In the frequency intervals where only one effective parameter is negative, either the mass density or the Young's modulus (shear modulus), the propagation of waves is forbidden. In contrast, if the structure is engineered to support frequency intervals where the doubly negative property occurs, *i.e.* simultaneously negative effective mass density and modulus, phenomena not present in nature may arise, as for instance, the negative refraction or the cloaking effect. In the past decade, a couple of configurations allowing for the double negativity have been reported in theoretical[10–19] as well as in experimental[11,17] works. However, most of these studies were focusing on bulk waves whereas the control over other types of elastic waves, such as the Lamb waves, is a prerequisite to the development of planar double-negative elastic metamaterials.[20] Among the most suitable candidates in that respect are probably the pillared metamaterials[21–26] which could be described as phononic stubbed plates constructed by depositing cylindrical dots on a thin homogeneous membrane.[27] Their peculiar elastic properties ensue from the vibration of the pillars at resonance that couples with the wave propagating in the plate. Although to different extents, three kinds of resonances may be involved in the dynamic properties of these systems, namely the bending, the compressional and the rotational modes.[26] In contrast to the bending and the compressional resonances, the combination of which has been reported to turn mass density negative in single-sided pillared metamaterials,[31] less attention has been paid to the rotational resonance to date. Interestingly, it has been theoretically demonstrated that rotational resonance of a core mass in a mass-spring system can lead to negative effective stiffness[12,16,28] and designs involving rotational inertia have been proposed to demonstrate both numerically and experimentally the occurrence of the double negativity[15,17,29] which in turn widen the scope of applications. Remembering that the double negativity can be achieved either by combining two different substructures, each supporting a different resonant mode[3,11,13] or by constructing a single structure where two resonances occur at a single frequency,[12,15–17,26] we propose in this letter a new path to achieve the doubly negative property that consists in combining the bending, compressional and rotational modes into one pillared system.

We first describe and analyze the dynamic behavior of a single-phase asymmetric double-sided pillared metamaterial (DPM) whose unit cell is shown in Fig. 1(a). We show that the negative effective mass density (NMD) results from the combination of the bending and compressional resonances of one pillar whereas the rotational resonance of the other pillar leads to the negative effective shear modulus (NSM). Lastly, numerical simulations of the zero-index refraction are carried out to put into evidence the doubly negative property.

Two distinct pillars (labelled as A and B respectively) are concentrically assembled over a thin matrix plate. Their dimensions were laid down for the resonances to occur in the MHz range: the diameter and height of pillar A (resp. pillar B) were $d_A = 80\mu m$ ($d_B = 110\mu m$) and $h_A = 200\mu m$ ($h_B = 130\mu m$); the lattice constant and the thickness of the plate were $a = 200\mu m$ and $e = 100\mu m$ respectively. Both the matrix plate and pillars were made of steel whose Young's modulus, Poisson's ratio, and mass density are $E = 200GPa$, $v = 0.3$ and $\rho = 7850 kg.m^{-3}$ respectively. Because of the asymmetry with respect to the mid-plane of the matrix plate, it can be anticipated that the symmetric and antisymmetric Lamb waves cannot be decoupled. Before investigating the double-sided pillared system, we have studied separately the two single-sided pillared metamaterials depicted in Figs. 1(b) and 1(c). Each of them was built with pillar A or pillar B erected in the center of a square unit cell of side $a$, on a plate having a thickness $e$; we refer hereafter to these systems as SPMA and SPMB. Their band structures computed using a finite element method are displayed in Figs. 1(e) and 1(f) respectively. The band structure of SPMA comprises a low frequency band gap that opens up in between 5.19MHz and 5.47MHz. The flatness of the dispersion curves around the lower limit of this band gap suggests that it results from a local resonance of the pillar. This is further evidenced by the eigenmodes at point M of the first irreducible Brillouin zone (BZ), labelled as C and D in Figs. 1(e) and 2(a). The result displayed in Fig. 2(b) unambiguously shows that these eigenmodes are the second-order bending resonance and the first-order compressional resonance of the pillar. The next step was to evaluate the 3×3 dynamic effective mass density matrix $[\rho]_{eff}$. The method consists of applying an external displacement field **U** on the four lateral boundaries of the unit cell while leaving the other two faces free. The induced force **F** is then derived by evaluating the stress average over the four boundaries.[10,15,30] In the harmonic regime at frequency $\omega/2\pi$, **F** and **U** are related by $\mathbf{F} = -\omega^2 V [\rho]_{eff} \mathbf{U}$, where $V$ denotes the volume of the unit cell. Both the normalized components $\rho_{11}$ ($\rho_{22} = \rho_{11}$ because of the square symmetry of the unit cell) and $\rho_{33}$ against the excitation frequency are shown in Fig. 1(g). Both components turn negative from 5.32MHz to 5.49MHz, in good agreement with the stop band shown in Fig. 1(e) that goes from 5.19MHz to 5.47MHz. The small discrepancy of about 2.5% at the lower edge of the band gap may be readily ascribed to the phase change across the unit cell, not accounted for in the calculation since this numerical method is only valid in the long wavelength limit. It can be stated from this analysis that the low frequency band gap relates to NMD caused by the combination of the bending and compressional resonances of pillar A.

Regarding the band structure of SPMB displayed in Fig. 1(f), no complete band gap arises in the investigated frequency range from 0MHz to 7MHz. The eigenmodes at points Γ and M, labelled as E and F in Figs. 1(f) and 2(a), show that pillar B undergoes an alternative rotational motion around its center axis. One may suspect that this rotational motion can couple with the local shear deformation of the matrix plate allowing in turn the effective shear modulus $\mu_{eff}$ to turn negative. To verify this assumption, we have calculated $\mu_{eff}$ using the numerical method described in Refs.[10,29,31]. In the calculation, we have considered a simple shear strain field applied along two parallel lateral boundaries. This sets the local displacement field and excites the rotational vibration of pillar B. The behavior of $\mu_{eff}$ against the excitation frequency is then deduced from the equivalence between the energy of the induced force vector on the lateral boundaries of the unit cell and the strain energy of the effective medium. The relationship between them can be expressed as $\sum_{\partial V} \mathbf{F} \cdot \mathbf{U} = \frac{1}{2}\mu_{eff}\gamma^2 V$, where $\partial V$ stands for the lateral boundaries of the unit cell and $\gamma$ represents the applied simple shear strain. The result displayed in Fig. 1(h) shows that $\mu_{eff}$ is negative from 5.29MHz to 5.36MHz, in very good agreement with the frequency interval in between points F (5.29MHz) and E (5.35MHz). Therefore, a locally resonant band gap should be expected in this region. However, because of the dispersion of the compressional resonance labelled as G in Fig. 1(f) no complete band gap opens up in this interval. It should be pointed out that pillar B was specifically designed in such a way that its rotational resonance falls inside the frequency range of NMD achieved in SPMA. In this case, the double negativity can be expected when combining both pillar A and pillar B to form DPM. In this merged structure, NMD would result from the bending and compressional resonances of pillar A whereas NSM would ensue from the rotational resonance of pillar B.

To validate this approach, we have computed the band structure of DPM. As expected, an isolated negative-slope branch, highlighted in red in Fig. 1(d), appears in between 5.28MHz and 5.35MHz. Additionally, we show in Fig. 2(b) the displacement field at some characteristic points, labelled from C' to G' in Figs. 1(d) and 2(a). Comparing the band structures of these three pillared metamaterials allows understanding the formation of this branch. They are drawn in Fig. 2(a) where the black, red and blue dotted lines represent the dispersion along ΓM direction of DPM, SPMA and SPMB respectively. At point M of the BZ, both the bending (point C) and the compressional modes (point D) slightly shift to points C' and D' upon attachment of pillar B to the plate. For both these resonances, the displacement fields of DPM displayed in Fig. 2(b) show that the deformation of pillar B is very small at the compressional resonance or even null at the bending resonance. This suggests that pillar B acts as an inert mass attached to the plate that simply shifts the resonant frequencies of pillar A. Accordingly, the frequency interval of NMD generated by resonances C' and D' of pillar A in DPM also shifts and appears now in between 5.21MHz and 5.48MHz instead of 5.19MHz to 5.47MHz in SPMA, but the overall mechanism leading to NMD is the same for both structures.

The situation is totally different when comparing the band structures of DPM and SPMB. In this case appending pillar A to SPMB does not summarize into a simple shift of the resonant frequency of pillar B: at point M, the compressional resonance labelled as G' in Fig. 2(a) affects both pillars in DPM (see Fig. 2(b) panel G') and therefore the shift from G (compressional resonance of SPMB) to G' cannot be ascribed to an inert mass attached to the plate like before. At the same time, the eigenfrequencies at points labelled as E and F in Figs. 1(f) and 2(a) remain unchanged because there is no coupling between the rotational vibration of pillar B and the bending and compressional vibrations of pillar A. For a sake of coherency in the notations, these points are labelled as E' and F' in Figs. 1(d) and 2(a). As mentioned before, the effective shear modulus turns negative in the frequency interval between these two points. Therefore, both NMD and NSM are achieved in this interval which perfectly explains the occurrence of the negative-slope propagative branch in Fig. 1(d). More generally, the preceding analysis demonstrates that the double negativity can be obtained if the bending, compressional and rotational resonances of the pillared system are well designed to occur within the same frequency interval but it says nothing on how to excite these resonances.

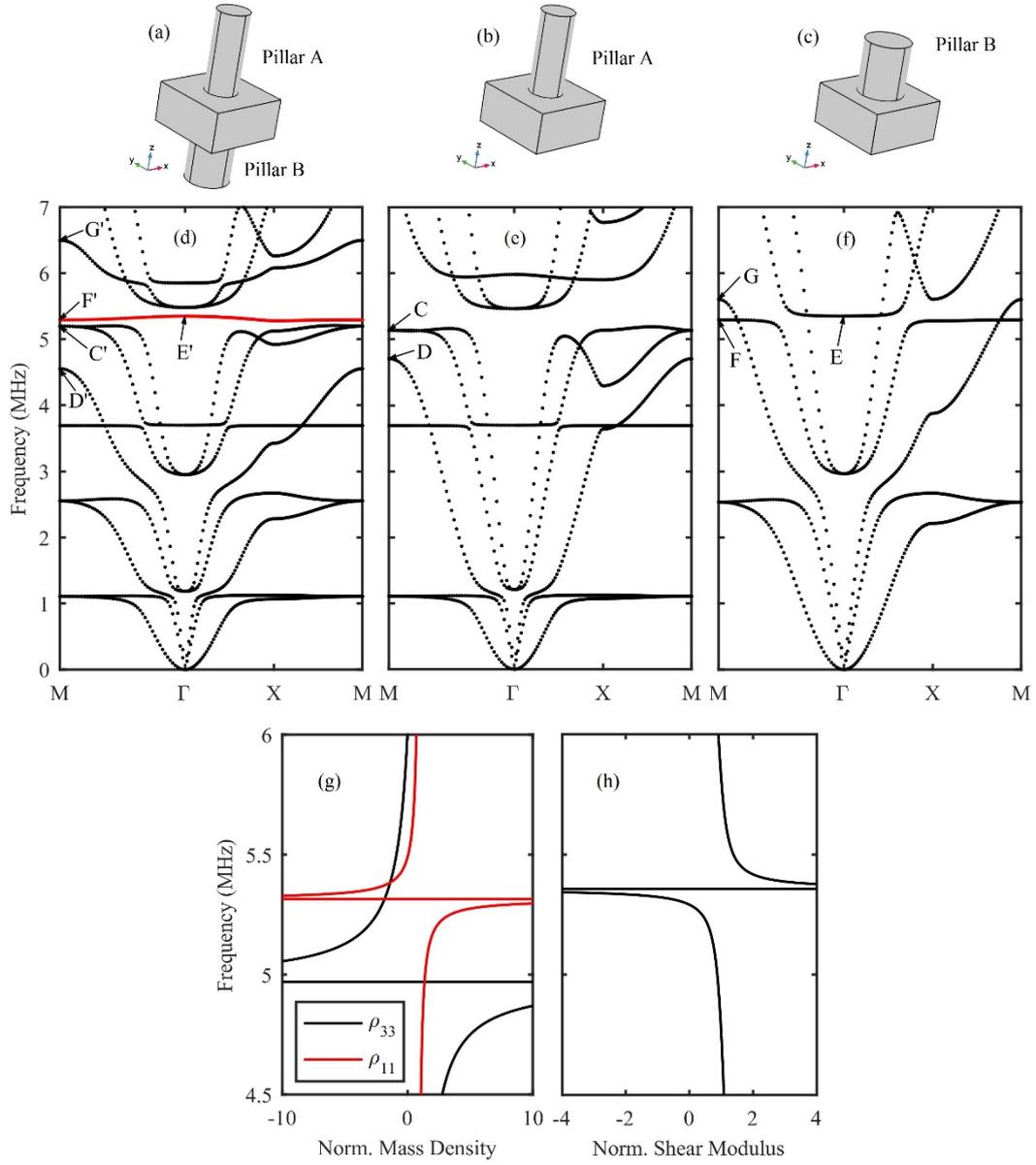

**Figure 1:** (a)-(c) Representative square lattice unit cells of DPM, SPMA and SPMB respectively and (d)-(f) their corresponding band structures. (g) Normalized effective mass density components $\rho_{33}$ (black line) and $\rho_{11}$ (red line) of SPMA. (h) Normalized effective shear modulus of SPMB.

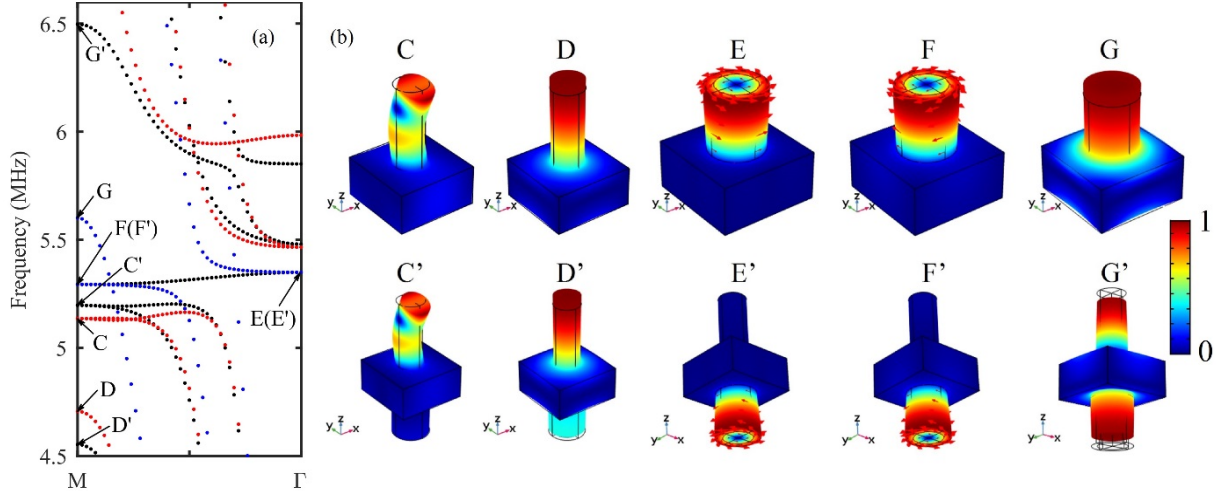

**Figure 2:** (a) Comparison of the band structures of DPM (black dotted lines), SPMA (red dotted lines) and SPMB (blue dotted lines) along ΓM direction. (b) Normalized total displacement and deformation of the unit cell corresponding to the points indicated in panel (a) and Figs. 1(d)-1(f).

One might take advantage of the in-plane polarization of a SH Lamb wave to trigger the rotational vibration of the pillar and actually we have verified that an incident wave with the frequency inside the double-negative branch can propagate across the metamaterial (not shown here). However, the other two types of Lamb wave waves, *i.e.* the symmetric and the antisymmetric Lamb modes, are polarized in the sagittal plane and therefore they cannot excite the rotational resonance because of the mirror symmetry in the unit cell. To overcome this difficulty, chirality may be introduced in the pillar so that the waves propagating in the plate can create an asymmetric deformation. Both the cross section and the side view of pillar B fulfilling this requirement is shown in Fig. 3(a). Eight flanks equally spaced in azimuth with a length $l = 60$μm and a width $w = 10$μm are inserted along a solid cylinder with a diameter $d = 100$μm. Pillar B is formed by stretching the cross section along negative $z$ direction with height $h = 105$μm and a twist angle $\theta = 45°$ in anti-clockwise direction as shown in the bottom panel of Fig. 3(a). The corresponding band structure along ΓX direction is displayed as red lines in Fig. 3(b). The double-negative branch goes from 5.37MHz to 5.41MHz. To illustrate the efficiency of chirality in exciting the rotational vibration, the transmission spectrum of an antisymmetric Lamb wave impinging at normal incidence a structure made of nine unit cells along *x*-axis and infinite along the *y*-direction is displayed as red lines in Fig. 3(c). For comparison, the black solid lines in Fig. 3(b) and 3(c) correspond to DPM shown in Fig. 1(a). In the absence of chirality, the transmission coefficient at a frequency in the double-negative branch is null since the rotational resonance is not excited with such a design. In contrast, the chiral pillars allow for a transmission coefficient of about 0.25 thanks to the combination of the bending and compressional vibrations of pillar A and the rotational vibration of pillar B, which is the key point for the occurrence of the doubly negative property.

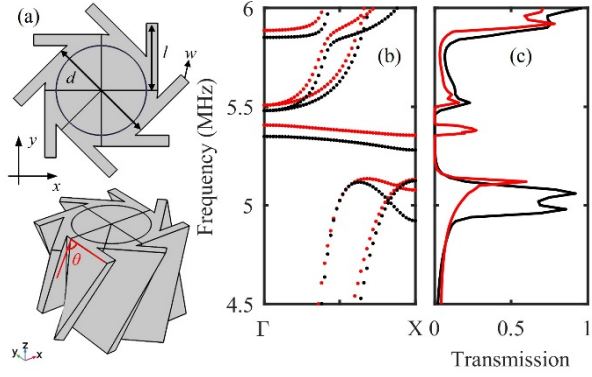

**Figure 3**: (a) Representative profile of the chiral pillar B. (b) Band structure along ΓX direction of DPM involving the chiral pillar B (red lines) or without chirality (black lines). (c) Transmission spectrum of an antisymmetric Lamb wave impinging at normal incidence on the phononic crystal with (red line) and without (black line) the chiral pillar.

One of the most amazing properties that ensues from the double negativity is the cloaking effect. At a frequency in the negative-slope branch, this effect results from the fact that both phase velocity and wavelength become infinite and in turn we find the case of a zero refractive index material.[10,20,32–34] We have investigated this effect at the frequency of 5.4MHz, where the effective shear modulus tends toward infinity. The FEM model is shown in the top panel of Fig. 4. It consists of 132 unit cells and features a $7a\times3a$ rectangular void in its center. A zero-order antisymmetric Lamb wave is excited at a distance of 1mm from the left edge of the metamaterial and perfectly match layers are implemented on each side of the sample to eliminate any reflection from the boundaries. Periodic boundary conditions are applied on the other two edges. The out-of-plane component of the displacement field at 5.4MHz is displayed in the middle panel in Fig. 4. It can be seen that the wave front keeps plane upon transmission through the sample, except around the void where scattering effects are observable. As a consequence of the infinite effective shear modulus and finite effective mass density in the metamaterial, the phase velocity gets nearly infinite and there is no phase change of the antisymmetric Lamb wave propagating in the metamaterial, allowing for a cloaking effect in this system. In contrast, when the working frequency is tuned to 6MHz, *i.e.* a frequency where the effective shear modulus is positive, the incident antisymmetric Lamb wave undergoes strong scattering on the void, giving rise to the distorted wave front observable in the bottom panel of Fig. 4. This simple analysis of the transmission through the pillared system unambiguously shows that the shielding of substructures at specific frequencies may be achieved with this geometry.

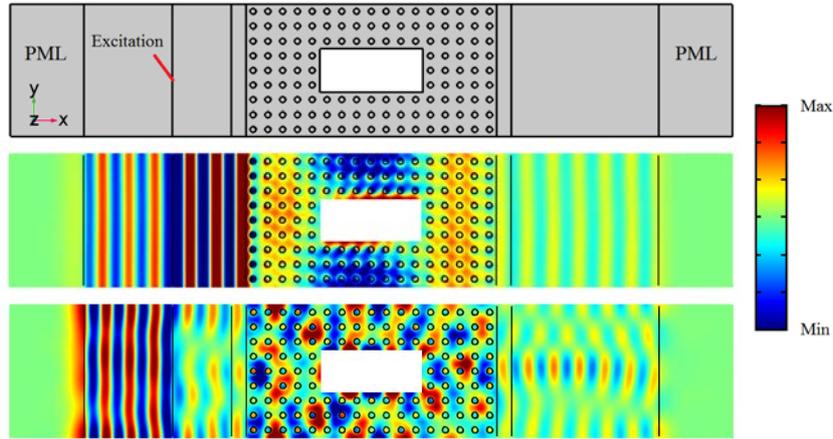

**Figure 4:** FEM model implemented to verify the cloaking effect (top panel); out-of-plane component of the displacement field upon antisymmetric excitation at frequencies 5.4MHz (middle panel) and 6MHz (bottom panel).

To conclude, we have realized the doubly negative property in an asymmetric double-sided pillared metamaterial. The mechanism responsible for the negative effective mass density is described as being the combination of the bending and compressional resonances of one pillar, whereas the negative effective shear modulus results from the rotational resonance of the other. This design contributes to broaden the field of applications of the pillared metamaterials that includes the negative refraction and over-diffraction-limit imaging of Lamb waves.